\documentclass[twocolumn,showpacs,prl,floatfix]{revtex4}
\usepackage{amssymb}
\usepackage{amsmath}
\usepackage{amstext}
\usepackage{latexsym}
\usepackage{graphics}
\usepackage{graphicx}
\usepackage{amsfonts}

\def\bk{\mathbf{k}}

\begin{document}

\title{Surface plasmon modes and the Casimir energy}
\author{F. Intravaia}
\affiliation{Laboratoire Kastler-Brossel UPMC/ENS/CNRS case 74, Campus Jussieu, F75252
Paris Cedex 05}
\author{A. Lambrecht}
\affiliation{Laboratoire Kastler-Brossel UPMC/ENS/CNRS case 74, Campus Jussieu, F75252
Paris Cedex 05}


\begin{abstract}
We show the influence of surface plasmons on the Casimir effect between two plane parallel metallic mirrors at arbitrary distances. 
Using the plasma model to describe the optical response of the
metal, we express the Casimir energy as a sum of contributions associated with
evanescent surface plasmon modes and propagative cavity modes. In contrast to naive expectations, the plasmonic modes contribution is essential at all distances in order to ensure the correct result for the Casimir energy. One of the two plasmonic modes gives rise to a repulsive contribution, balancing out the attractive contributions from propagating cavity modes, while both contributions taken separately are much larger than the actual value of the Casimir energy. This also suggests possibilities to tailor the sign of the Casimir force via surface plasmons.
\end{abstract}

\date{\today}
\maketitle

When H. Casimir first predicted the existence of a force between neutral
mirrors in vacuum \cite{Casimir48}, he considered two plane parallel perfect
reflectors and found an interaction energy $E_{\text{Cas}}$ depending only on geometrical parameters, the mirrors distance $L$ and surface $A\gg L^{2}$, and two fundamental
constants, the speed of light $c$ and Planck constant $\hbar $ 
\begin{equation}
E_{\text{Cas}}=-\frac{\hbar c\pi^2 A}{720L^{3}}. \label{CasimirForce}
\end{equation}%
The signs have been chosen to fit the thermodynamical convention with
the minus sign of the energy $E_{\text{Cas}}$
corresponding to a binding energy. The Casimir energy for perfect mirrors is usually
obtained by summing the zero-point energies $\frac{%
\hbar \omega }{2}$ of the cavity eigenmodes, substracting the
result  for finite and infinite separation, and 
extracting the regular expression (\ref{CasimirForce}) by inserting a formal
high-energy cutoff and using the Euler-McLaurin formula \cite{qft}.

In his seminal paper \cite{Casimir48}, Casimir noticed that the energy
should be a finite expression, without the need of any regularization,
provided one takes into account the high frequency transparency of  real mirrors. The idea was implemented by Lifshitz
who calculated the Casimir energy for mirrors characterized by
dielectric functions \cite{Lifshitz56}. For metallic mirrors he recovered expression (\ref
{CasimirForce}) for separations $L$ much larger than the
plasma wavelength $\lambda _{\mathrm{p}}$ associated with the metal, as metals are very good reflectors at frequencies
much smaller than the plasma frequency $\omega _{\mathrm{p}}$. At shorter
separations in contrast, the Casimir effect probes the optical response of
metals at frequencies where they are poor reflectors and the Casimir energy is reduced
with respect to (\ref{CasimirForce}). This reduction has been
studied in great detail recently (\cite{Jaekel91,GenetPRA03} and references
therein) since it plays a central role in the comparison of theoretical predictions (\cite{compar} and references therein) with experimental results 
\cite{expts}.

In the limit of small separations $L \ll\lambda _{\mathrm{p}}$, the Casimir effect has another interpretation establishing a bridge between quantum field theory of vacuum fluctuations
and condensed matter theory of forces between two metallic bulks. It can
indeed be understood as resulting from the Coulomb interaction between 
surface plasmons, that is the collective electron excitations propagating
on the interface between each bulk and the intracavity vacuum \cite
{plasmon,Barton79,Schram73}. The corresponding field modes are evanescent waves and have an imaginary longitudinal wavevector. We will call them plasmonic modes at arbitrary distances as they coincide with the surface plasmon modes at small distances. Plasmonic modes have to be seen in contrast to ordinary propagating cavity modes, which have a real longitudinal wavevector. For simplicity we will call those in the following photonic modes. Photonic modes are usually considered in quantum field theory of the Casimir effect \cite
{qft} and are thought to determine the Casimir effect at large distances where the
mirrors can be treated as perfect reflectors. At short distances, plasmonic modes are known to dominate the interaction \cite{GenetAFLB03,Henkel03}. 

The purpose of the present letter is to show the singular behavior of one of the two plasmonic modes, which gives rise to a repulsive contribution to the Casimir energy at all distances, ensuring in this way that the correct value for the Casimir energy is recovered, in particular the ideal Casimir energy at large distances. Plasmonic modes have therefore a much greater importance than usually appreciated. To show this, we will use the decomposition of the Casimir energy
as a sum of zero-point energies $\frac{\hbar \omega }{2}$ over the whole set
of modes of the cavity with its two mirrors described by a
plasma model. This set contains plasmonic as well as photonic modes. As expected from \cite{GenetAFLB03,Henkel03}, the contributions of plasmonic modes will be
found to dominate the Casimir effect for small separations corresponding to Coulomb interaction between surface plasmons.
But, contrary to naive expectations, they do not vanish for large separations. For distances larger than about $\lambda_{\mathrm{p}}/4\pi$ ($\sim$10nm for typical metals) they even give rise to a contribution having simultaneously
a negative sign and a too large magnitude with respect to the Casimir formula
(\ref{CasimirForce}). The repulsive character can be attributed to one of the two plasmonic modes. The photonic modes as well as the second plasmonic mode give rise to an attractive contribution much larger than (\ref{CasimirForce}). It is therefore the repulsive contribution of a single plasmonic mode which renders the total plasmonic mode contribution to the Casimir energy repulsive outside the short distance limit while assuring at the same time that the sum over all modes reproduces (\ref{CasimirForce}) at large distances.  
This repulsive character may open interesting possibilities
to tailor surface plasmons via nanostructuration of metallic surfaces in order to change the sign of the total Casimir force. 

In this letter, we restrict our attention to the situation of two
infinitely large plane mirrors at zero temperature so that the
only modification of Casimir formula (\ref{CasimirForce}) is due to the
metals finite conductivity. This modification is
calculated by evaluating the radiation pressure of vacuum fields upon the
two mirrors \cite{GenetPRA03} 
\begin{eqnarray}
E &=&-\sum_{\epsilon}\sum_{\mathbf{k}}\sum_{\omega }\frac{i\hbar}{2} \ln({1-r_{\mathbf{k}}^{\epsilon}[\omega
]^{2}e^{2ik_{z}L}}) + c.c.   \label{realForce} \\
\sum_{\mathbf{k}} &\equiv &A\int \frac{\mathrm{d}^{2}\mathbf{k}}{4\pi ^{2}}%
\quad ,\quad \sum_{\omega }\equiv \int_{0}^{\infty }\frac{\mathrm{d}\omega }{%
2\pi }.  \notag
\end{eqnarray}%
The energy $E$ is obtained by summing over 
polarization $\epsilon$=(TE,TM), transverse wavevector $\mathbf{k}\equiv \left(
k_{x},k_{y}\right) $ (with $z$ the longitudinal axis of the cavity) and 
frequency $\omega $; $k_{z}$ is the longitudinal wavevector associated with the mode.
The reflection amplitudes $r_{\mathbf{k}}^{\epsilon}$, here supposed to be the same
for both mirrors are causal retarded functions obeying high-frequency
transparency.

We now calculate the Casimir energy as a sum over
the cavity modes using the plasma model for the mirrors dielectric function 
\begin{equation}
\varepsilon \lbrack \omega ]=1-\frac{\omega _{%
\mathrm{p}}^{2}}{\omega ^{2}}
\end{equation}
with $\omega _{\mathrm{p}}$ the plasma frequency and $\lambda _{\mathrm{p}}=\frac{2\pi c}{\omega _{\mathrm{p}}}$ the plasma wavelength, of the order of 100nm for metals used in experiments \cite{Lambrecht00}.
In this case the zeros of the argument of the integrand in (\ref{realForce}) 
lie on the real axis. In
fact, they have to be pushed slightly below this axis by introducing a
vanishing dissipation parameter in order to avoid any ambiguity in 
expression (2) \cite{GenetPRA03}. We may then rewrite (\ref%
{realForce}) as a sum over the solutions $\left[ \omega _{\mathbf{k}}^{\epsilon}%
\right] _{m}$ of the equation labelled by an integer index $m$ 
\begin{equation}
r_{\mathbf{k}}^{\epsilon}[\omega ]^{2}e^{2ik_{z}L}=1.  \label{solModes}
\end{equation}
Simple algebraic manipulations exploiting residues theorem and complex integration techniques \cite{Schram73} then lead to the Casimir energy
expressed as sums over these modes
\begin{eqnarray}
E &=& \sum_{\epsilon,\mathbf{k}}\left[\sum_{m}^{\prime}\frac{\hbar \left[
\omega _{\mathbf{k}}^{\epsilon}\right]_{m}}{2}\right]_{L\rightarrow\infty}^{L}.\label{diff}
\end{eqnarray}%
The prime in the sum over $m$ signifies as usually that the term $m=0$ has to be multiplied by 1/2. The sum over the modes is to be understood as a regularized quantity as it involves infinite quantities. This result is well known for perfect mirrors and is not changed by the choice of the plasma model for the mirrors reflection coefficients. The upper expression contains as limiting cases at large distances the Casimir expression with perfect mirrors and at short distances the expression in terms of surface plasmon resonances. For arbitrary distances, photonic modes as well as plasmonic modes are important. 

\begin{figure}[tbp]
\centering\includegraphics[height=5.5cm]{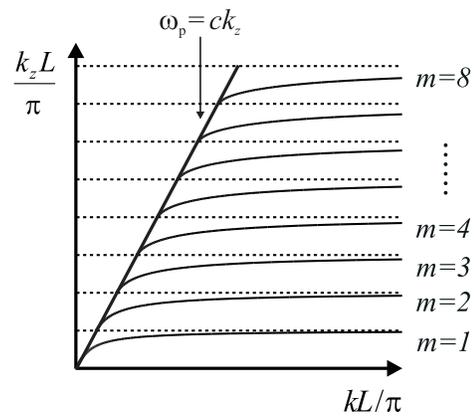}
\caption{Mode plot of the first photonic TE modes ($m=1,2,...8$) with the plasma model for $ck=0.5\omega_\text{p}$. Modes are presented through their longitudinal wavevector as a function of $kL/\pi$. The dotted lines correspond to the cavity modes with perfect mirrors.}
\label{modes2}
\end{figure}

We will now discuss the structure of TE and TM modes inside the cavity formed by the two mirrors. The different modes have been obtained by writing explicitly all solutions of (\ref{solModes}), using the standard expressions for the reflection coefficients. Figure \ref{modes2} shows the phase shift acquired by the TE modes  through the influence of imperfect reflection. They are represented through their longitudinal wavevector as a function of $kL$. The TE polarization admits only photonic modes which can be written under
the standard form $k_{z}L=m\pi -\delta $, where the integer $m=1,2\ldots \infty $ is the order of the cavity mode and $\delta $
the phase shift of the mode on a mirror.
Perfect mirrors lead to cavity modes plotted as
dotted lines corresponding to $\delta ^{\mathrm{TE}}=0$. 
With the plasma model, the photonic modes are
displaced compared to the perfect cavity modes as a direct consequence of the phase shift $\delta $ acquired by
vacuum fields upon reflection. The limit of perfect reflection corresponds to the large distances limit. The high frequency transparency of metallic mirrors imposes an upper bound
to their longitudinal wavevector $ck_{z}<\omega _{\mathrm{p}}$, where all photonic modes coincide. 

\begin{figure}[tbp]
\centering\includegraphics[height=5cm]{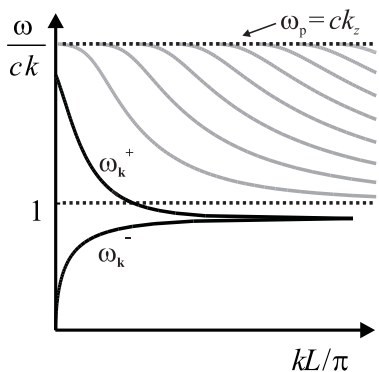}
\caption{Mode plot of the two plasmonic modes $\omega_\bk^-$ and $\omega_\bk^+$ (black) in the sector $\omega < ck$ and of photonic modes (gray) in the sector $\omega > ck$ for $ck=0.5\omega_\text{p}$. Modes are presented  through their frequency as a function of $kL/\pi$.}
\label{modes}
\end{figure}

For TM polarization,
similar photonic modes are obtained labelled also by a positive
integer $m$. They are accompanied by two additional modes, which we label $\left[ \omega _{\mathbf{k}}^{\mathrm{pl}}\right] _{\pm }$ as they 
tend to the frequencies of surface plasmon modes \cite{Barton79} in the limit of
small distances. These plasmonic modes are shown as solid black lines in Figure \ref{modes}, while photonic modes correspond to gray lines. In order to make the plasmonic modes with their imaginary wavevector visible, the modes are now represented through their frequency as a function of $kL$. Plasmonic  and photonic modes lie respectively in the sector  $\omega < ck$ and  $\omega > ck$.
In the limit of infinite mirrors separation, the plasmonic modes are given by the usual dispersion relation for the surface plasmons in a metallic bulk \cite{Barton79} 
\begin{equation}
\left[ \omega _{\mathbf{k}}^{\rm pl}\right]_{\pm} \underrightarrow{L\rightarrow\infty}\quad \frac{\omega_{\mathrm p}^2+2|\mathbf{k}|^2-\sqrt{\omega_{\mathrm p}^4+4|\mathbf{k}|^4}}{2}. \label{wsp} 
\end{equation}
For the photonic modes the phase shift $\delta$ tends towards zero for infinite distances where they obey the dispersion relation for perfect mirrors $\left[ \omega _{\mathbf{k}}^{\epsilon}\right]_{m}=\sqrt{\left\vert \mathbf{k}\right\vert ^{2}+k_{z}^{2}}$ with the longitudinal wavevector $k_z=m\pi/L$. For $L\rightarrow \infty$, the sum over $m$ in (\ref{diff}) becomes a continuous integral and the mode contribution of photonic modes corresponds to the one of free field vacuum which is substracted from the contribution at finite distances.

Let us now discuss in more detail the behavior of the two plasmonic modes. $\omega_\bk^-$  is restricted to the plasmonic mode sector, while $\omega_\bk^+$ lies in the plasmonic mode sector for large distances, but crosses the barrier $\omega = ck$ and dies in the photonic mode sector for $kL/\pi \rightarrow 0$. In the present calculation, the whole mode was attributed to the plasmonic mode contribution as its frequency tends to the surface plasmon contribution at short distances. The qualitative results do not change if the part of the mode lying in the photonic modes sector is attributed to the photonic modes contribution. 

Obviously, when decreasing the distance $L$ the plasmonic mode $\omega_\bk^+$  acquires a phase shift with the same sign as the TM photonic modes below the plasma frequency. Its frequency at short distances is always larger than the one in the large distance limit. In contrast, the frequency of $\omega_\bk^-$  is decreased at short distances compared to long distances. 
When now performing the difference (\ref{diff}) of the contributions at finite and infinite distances, the Casimir energy contribution turns out to be negative for photonic modes, as the mode contribution in free vacuum ($L\rightarrow\infty)$ exceeds the one inside the cavity, in accordance with an attractive force. It is also negative for the plasmonic mode $\omega_\bk^-$. However, the difference is positive for the plasmonic mode $\omega_\bk^+$.
An immediate consequence is that the contribution of $\omega_\bk^+$  to the Casimir energy is repulsive.

\begin{figure}[tbp]
\centering\includegraphics[width=8.5cm]{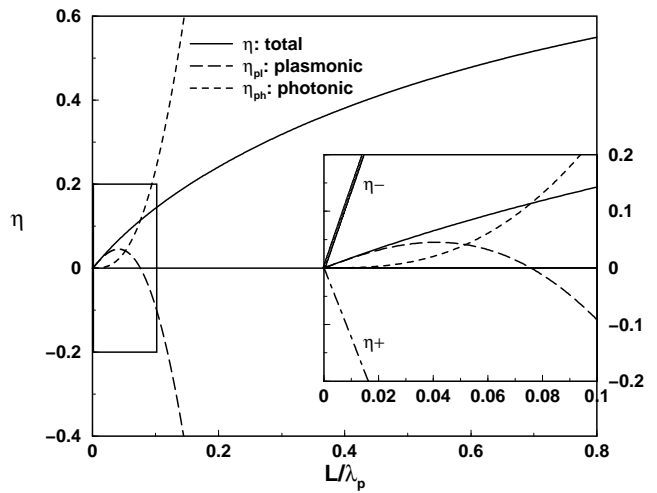}
\caption{Contributions to Casimir energy normalized to (1) of photonic modes (dotted) and plasmonic modes (dashed) to the total Casimir energy (solid line) as functions of $L/\protect\lambda _{\mathrm{p}}$. The inlet shows the separate contributions of $\omega_\bk^-$ and $\omega_\bk^+$.}
\label{beha}
\end{figure}

To asses quantitatively the effect of the plasmonic modes to the Casimir energy, we have computed separately the energies associated with photonic modes $\left[ \omega _{\mathbf{k}}^{p}\right]
_{m}$ and
plasmonic modes $\left[ \omega _{\mathbf{k}}^{\mathrm{pl}}\right]
_{\pm }$. 
All energies in the following will be presented as a reduction factor $\eta$
\cite{Lambrecht00} 
\begin{equation}
E=\eta E_{\text{Cas}}.  \label{realEnergy}
\end{equation}
As the ideal Casimir energy is negative corresponding to attraction, positive and negative  reduction factors mean respectively attractive or repulsive interaction. The reduction factor due to imperfect reflection described with the plasma model is shown as a solid line in Figure \ref{beha} as a function of the ratio $L/\lambda _{\mathrm{p}}$. 
We also introduce reduction factors corresponding to contributions of the different modes to the Casimir energy   
\begin{equation}
\eta _{\text{ph}}=E_{\text{ph}}/E_{\text{Cas}} \quad  \quad \eta _{\text{pl}}=E_{\text{pl}}/E_{\text{Cas}}.
\notag
\end{equation}
Their sum corresponds to the whole Casimir energy $\eta =\eta _{\text{ph}}+\eta _{\text{pl}}$.
The contribution $\eta _{\text{pl}}$ of plasmonic modes (dashed line)
dominates at short distances $L\ll \lambda _{\mathrm{p}}$, which confirms the interpretation of the
Casimir effect as resulting in this regime from the Coulomb interaction of
surface plasmons. There, a simple
expression may be given for the reduction factor \cite{GenetPRA00,GenetAFLB03} 
\begin{equation}
\eta \underset{L\ll \lambda _{\mathrm{p}}}{\simeq }\frac{3\alpha }{2}\frac{L%
}{\lambda _{\mathrm{p}}}\quad ,\quad \alpha \simeq 1.193.
\label{shortSeparations}
\end{equation}%
The power law dependence of $E$ then goes from $L^{-3}$ at large distances
to $L^{-2}\lambda _{\mathrm{p}}^{-1}$ at short distances \cite{Lifshitz56}. 
The contribution of photonic modes $\eta _{\text{ph}}$
scales as $\left( L/\lambda _{\mathrm{p}}\right) ^{4}$ and its contribution
may be neglected at the 1\% level up to $L/\lambda _{\mathrm{p}}\sim 0.2$.
At larger distances, $\eta _{\text{ph}}$ increases while $\eta _{\text{pl}}$ 
becomes negative at a distance of the order $\lambda _{\mathrm{p}}/4\pi$. 
 This clearly comes from the behavior of $\omega_\bk^+$, shown in the inlet, which gives a repulsive contribution at all distances. 
For example, the photonic and plasmonic contribution to the Casimir energy at $\lambda_\text{p}/L\sim 1$ are both about 36 times larger than the total Casimir energy between metallic mirrors. They are of opposite sign while the photonic contribution slightly dominates. 
For large separations $L/\lambda _{\mathrm{p}%
}\gg 1$, $\eta _{\text{ph}}$ tends to $+\infty $ while $\eta _{\text{pl}}$
tends to $-\infty $. The sum of the two contributions reproduces the known
value for $\eta $, which is positive and increasing
over all separations going from (\ref{shortSeparations}) to unity for large distances, where the Casimir formula (\ref{CasimirForce}) is recovered. This feature results from a compensation
between the large positive value of $\eta _{\text{ph}}$ and the large negative
value of $\eta _{\text{pl}}$. 
More precise asymptotic laws for the two contributions are
\begin{equation}
\eta _{\text{ph}}-1\underset{L\gg \lambda _{\mathrm{p}}}{\simeq }-\eta _{%
\text{pl}}\ \underset{L\gg \lambda _{\mathrm{p}}}{\simeq }\beta \sqrt{\frac{L%
}{\lambda _{\mathrm{p}}}}\quad ,\quad \beta \simeq 74.58.
\end{equation}
The behavior of the whole reduction factor is also recovered $
\eta \underset{L\gg \lambda _{\mathrm{p}}}{\simeq }1-2\lambda _{\mathrm{p}
}/\left( \pi L\right) $. 

These results clearly show the crucial importance of the surface plasmon contribution, not only for short distances where it
dominates the Casimir effect but also for long distances. For metallic mirrors the existence of surface plasmons are not an additional correction to the Casimir effect, but inherent to it. A single plasmonic mode $\omega_\bk^+$ ensures consistency with the Casimir energy between metallic mirrors at intermediate distances and with the Casimir formula (1) for perfect mirrors. 
If we had calculated the Casimir effect by accounting only for the photonic   modes,
we would have found a result much too large. The photonic modes and one of the plasmonic modes  are displaced by the phase shifts which induce a systematical
deviation towards a larger magnitude of Casimir energy. The
discrepancy which would be obtained in this manner is only cured by the
contribution of the $\omega_\bk^+$ plasmonic mode. The whole Casimir 
energy turns out to be the result of a fine balance between 
the large attractive photonic contribution and the large repulsive plasmonic  contribution. As already known from discussions of
arbitrary dielectric mirrors \cite{GenetPRA03}, the outcome of this balance
keeps the sign of a binding energy. However, this result
relies heavily on the symmetry of the Casimir geometry with two plane
mirrors. One might thus hope changing this behavior by enhancing the
contribution of plasmonic modes, by changing the geometry, using for example the hole arrays used to
enhance the transmission of light through metallic structures \cite
{holearrays} or nanostructured metallic surfaces. This could then play a role in micro-electro-mechanical
systems (MEMS) in which the Casimir force is known to have a great influence 
\cite{mems}.

\begin{acknowledgments}
Many thanks are due to S. Reynaud, C. Genet, M.-T. Jaekel and P.A. Maia Neto for discussions.F.I. thanks the Foundation Angelo della Riccia for financial support.
\end{acknowledgments}

\end{document}